\documentclass[journal]{IEEEtran}
\usepackage{graphicx,url,cite}
\usepackage{amsmath,amsfonts}
\usepackage{amssymb}
\usepackage{multirow}
\usepackage{balance}

\def\R{\mathbb{R}}

\def\train{\texttt{train} }
\def\val{\texttt{val} }
\def\test{\texttt{test} }

\begin{document}
\title{Multichannel Attention Network for Analyzing Visual Behavior in Public Speaking}
\author{Rahul~Sharma, Tanaya~Guha and Gaurav~Sharma \vspace{0.3em}\\ 
IIT Kanpur \\ 
\texttt{\small\{rahus, tanaya, grv\}@iitk.ac.in}
}
\maketitle

\begin{abstract}
Public speaking is an important aspect of human communication and interaction. The majority of computational work on public speaking concentrates on analyzing the spoken content, and the verbal behavior of the speakers. While the success of public speaking largely depends on the content of the talk, and the verbal behavior, non-verbal (visual) cues, such as gestures and
physical appearance also play a significant role. This paper investigates the importance of visual cues by estimating their contribution towards predicting the popularity of a public lecture. For this purpose, we constructed a large database of more than $1800$ TED talk videos. As a measure of popularity of the TED talks, we leverage the corresponding (online) viewers' ratings from YouTube. Visual cues related to facial and physical appearance, facial expressions, and pose variations are extracted from the video frames using convolutional neural network (CNN) models. Thereafter, an attention-based long short-term memory (LSTM) network is proposed to predict the video popularity from the sequence of visual features. The proposed network achieves state-of-the-art prediction accuracy indicating that visual cues alone contain highly predictive information about the popularity of a talk. Furthermore, our network learns a human-like attention mechanism, which is particularly useful for interpretability, i.e.\ how attention varies with time, and across different visual cues by indicating their relative importance.
\end{abstract}


\IEEEpeerreviewmaketitle
\section{Introduction}
\label{sec:intro}
\IEEEPARstart{A}{nalysis} and modeling of human behavior are critical for human-centric systems to predict the
outcome of a social interaction, and to improve an interaction between humans or between human and
computer. Human behavior is expressed and perceived in terms of verbal, e.g.\ spoken dialogs, pitch,
intonation, and non-verbal cues, e.g.\ hand and body gestures, facial expressions, eye gaze,
appearance \cite{vinciarelli2009social}. These behavioral cues can be captured and processed to
predict the outcome of social interactions. Computational effort in this domain, though relatively
recent, is as diverse as analyzing couples' interaction \cite{xiao2015head},
to personality analysis \cite{batrinca2016multimodal}, to video blogging style analysis \cite{aran2014broadcasting,biel2013youtube}.
\begin{figure}[h]
\centering
\includegraphics[width=1.0\columnwidth ]{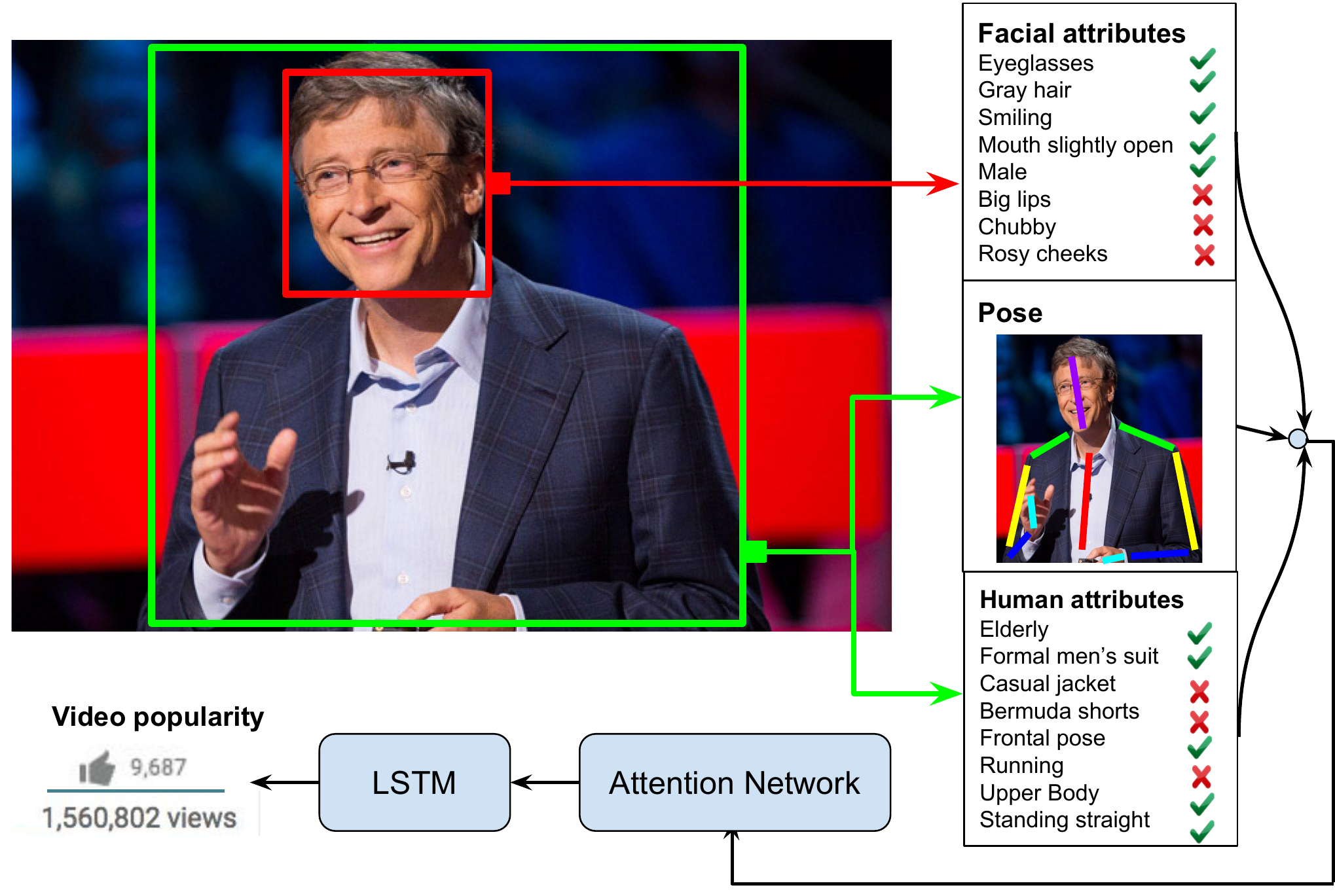}
\caption{The main idea of the proposed framework for predicting public lecture popularity from visual cues.}
\label{fig:teaser}
\end{figure}

Public speaking is an important aspect of human communication and interaction. A good speaker
is articulate, has convincing body language, and often, can significantly influence people
\cite{wesson2005communication}. While the success of public speaking largely
depends on the content of the talk, and the speaker's verbal behavior, non-verbal (visual) cues, such as gestures and
physical appearance also play a significant role \cite{riggio1986impression}. In this paper, we investigate the importance of visual cues within a predictive framework, i.e.\ the significance of visual cues is analyzed in terms of their ability to predict the effectiveness and success of public speaking.

In order to perform a controlled and objective study, we choose to analyze videos from the very popular
Technology, Entertainment, Design (TED)
seminar series. We constructed a database consisting of
$1864$ TED talk video, and collected relevant
statistics (number of \emph{views, likes, dislikes} and \emph{comments}) associated with each video
from YouTube. As an estimate of effectiveness or success of a TED talk, we propose to use the ratio of the number of \emph{likes} to that of the
\emph{views}. This is referred to as the \emph{video popularity} in the rest of the paper. 
 
We develop a computational framework that predicts the popularity of public speaking videos from visual cues. The main idea of the framework is summarized in Fig.~\ref{fig:teaser}. We hypothesize that the visual cues related to face, gesture, and physical appearance of a speaker together contribute to the popularity of a public lecture. We are thus
interested in a method for fusing multiple channels of different visual cues that can eventually
predict the video popularity. 

Motivated by the recent success of the long short-term memory (LSTM)
networks in sequence prediction tasks \cite{cho2015describing,yang2016stacked}, we base our work on them. A first approach to multichannel fusion within LSTMs would be to concatenate the channel encodings (pertaining to the individual visual cues) to
form a single monolithic feature. This approach serves as a competitive and contemporary
\emph{baseline} for our task. 

The simple concatenation of different channel encodings however is sub-optimal, since the features may lie in
distinct spaces with their respective different properties. To address this issue, we perform an alignment of the channels within the LSTM framework, and learn the alignment parameters for the current task along with all other parameters of the network. 
Further, we integrate attention mechanism into the framework. At every time step, we predict
which feature channel is the most relevant by learning the attention scores as a latent variable in the network. The proposed multichannel attention is novel, and different from the spatial attention networks, such as, stacked attention networks (SAN) \cite{yang2016stacked}, and we discuss this in more detail in Section \ref{sec:related}. Incorporating multichannel attention gives us the benefit of interpretability, as the attention scores on the different channels provide insights to the relative importance of the visual cues as time progresses.
Our full network architecture  
consists of three convolutional neural network (CNN) streams corresponding to different visual cues pertaining to physical appearance, gestures and facial expressions.
The information from these three channels is fused together (after learning an alignment of the channels) with the proposed multichannel attention-based LSTM network to predict video popularity.

The contributions of this work are as follows: (i) A large database containing videos from TED talks and their corresponding YouTube metadata is constructed to facilitate the study of public speaking in general. (ii) A novel architecture based on channel alignment, and multichannel attention LSTM is proposed for predicting video popularity from visual cues. This network enables better fusion of the visual cues, which yields state of the art results for our task. Furthermore, the network fosters interpretability and analysis of the visual cues, providing further insight to the significance of different visual cues in public speaking.

\section{Related Work}
\label{sec:related}
Human behavior in the context of public speaking has been studied extensively from the psychological and social perspectives. For example, psychological studies have investigated the influence of non-verbal cues \cite{riggio1986impression}, importance of using confidence cues (phrases that express speaker's confidence) in speech \cite{wesson2005communication}, the effect of physical distance \cite{albert1970physical} in speakers likability and persuasiveness, and the fear of public speaking \cite{pertaub2002experiment, daly1989self}. In the computational front however, work on automatic modeling, analysis and prediction of public speaking behavior is relatively limited. 

The existing literature of computational analysis of public speaking suggest the use of various modalities including speech, video, motion capture (MoCap), and even, manual annotation of behavioral activities. The majority of work involves analysis of speech and verbal behavior. Related work on speech include acoustic, prosodic and lexical analysis to discover vocal characteristics of a good speaker \cite{rosenberg2005acoustic, strangert2008makes}, and quantifying a speaker's attractiveness \cite{gonzalez2013perceptually}. A shift-invariant dictionary learning method was proposed to detect human-interpretable behavioral cues such as hand gestures, pose, and body movements from MoCap data \cite{tanveer2015unsupervised}. 
In another recent work, researchers attempted to automatically identify the nonverbal behavioral cues that are correlated with human experts' opinion of speaker performance \cite{wortwein2015multimodal, wortwein2015automatic}. An automatic performance evaluation was done using a database of $47$ people presenting in front of a virtual audience.
In a related work on job interviews, non-verbal behavioral cues were used to estimate a candidate's hirability \cite{nguyen2013multimodal}. More recently, a deep multimodal fusion architecture was proposed to predict persuasiveness of a speaker that indicates the influence a speaker has on the beliefs of an audience \cite{nojavanasghari2016deep}. This framework used video, audio and text descriptors to predict persuasiveness on a publicly available database containing more than $200$ videos. The descriptors used in their work consisted of standard acoustic and text features, and several hand crafted visual features.

Compared to the existing literature, the work presented in the
current paper studies public speaking at a much larger scale. Our approach is also completely
data-driven, and does not use any manual annotation for encoding behavioral cues. Our framework uses
the highly successful CNN architectures, namely the AlexNet \cite{krizhevsky2012imagenet}, and the
VGGNet \cite{simonyan2014very} for capturing visual cues. The modeling of sequential data is based on a
variant of the recurrent neural networks (RNN), called the LSTM \cite{hochreiter1997long}.
The attention-based framework proposed in our paper is
inspired by the success of the attention models used in various visual recognition tasks
\cite{adascan2017, yang2016stacked}. Perhaps the attention network the most
related to ours is the SAN \cite{yang2016stacked}. However, our method
differs from SAN in the following ways: (i) SAN considers \emph{spatial} attention for the task of
visual question answering, while our network computes \emph{temporal} attention across multiple
channels for video popularity prediction. (ii) SAN (and other attention networks) assumes the availability of information from
the entire data to compute attention, while in our proposed architecture, attention is
predicted at every time step, based on the information from the multiple channels in past frames, and those in the current frame.
\section{Database Collection}
\label{sec:database}

\begin{figure*}[t]
	\centering
	\includegraphics[width=1.0\textwidth, trim = 0cm 4cm 0cm 0cm, clip]{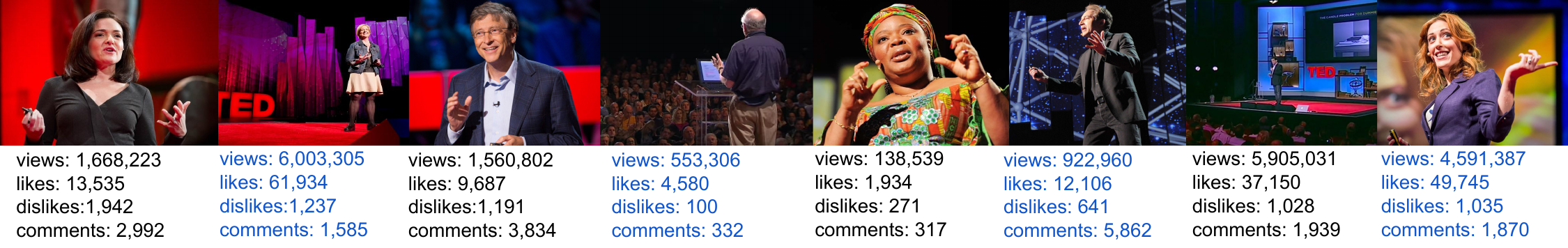}
	\caption{Sample frames from our TED1.8K database along with YouTube metadata.}
	\label{fig:data_sample}	
\end{figure*}
To facilitate the study of public speaking behavior, we constructed a large video database, namely the \textbf{TED1.8K} database. This database contains $1864$ TED talk videos, and their associated metadata collected from YouTube. The YouTube metadata that we collected for each video are - number of \emph{likes}, \emph{dislikes}, \emph{views} and \emph{comments}. Fig.~\ref{fig:data_sample} shows sample frames from our TED1.8K database, which demonstrates the huge variability, and the challenging nature of the database. Table \ref{tab:datasummary} presents a summary of the database.

\begin{figure}[t]
\centering
	\begin{minipage}{\linewidth}
		\includegraphics[width = 1.0\linewidth, trim = 0cm 15cm 0cm 0cm, clip]{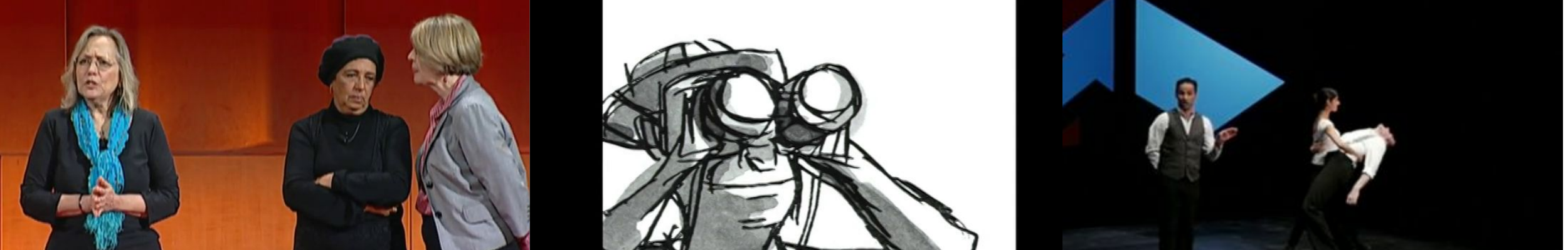}
		\caption{Sample frames from some of the discarded videos}
		\label{fig:rejected_vids}
	\end{minipage}		
\label{fig:likability}
\end{figure}

We first collected all the TED talk videos (initially, more than $2000$ videos) published until June 2016. We discarded videos for which (i) the YouTube `views' field was empty, or (ii) speaker body/face could not be detected in the majority of frames (for example, talks accompanied with dance performance; see Fig.~\ref{fig:rejected_vids} for examples). After this screening, $1864$ videos remained. 
We propose to estimate the popularity of the public lectures, $y$, as the ratio of its number of \emph{likes} to that of the \emph{views}.
\begin{align*}
	{\rm video\,\ popularity}\,\ y = \frac{{\rm number\,\ of\,\ }likes}{{\rm number\,\ of \,\ }views}
\end{align*}

\noindent 
These scores are normalized and mean-centered before feeding them to the regression network. 
The TED1.8K corpus provides certain advantages for studying human behavior in public speaking. Firstly, the video content is carefully created to have well defined audiovisual structure, with subtitles and transcripts of the talks. Therefore, the database offers opportunities for rich multimodal studies. Secondly, the TED talks are of diverse topics, and popular worldwide. Hence, the YouTube ratings are expected to come from viewers with varied demographics, age group, and social background, making the ratings rich and reliable. 
\begin{table}[tb]
    \centering
    \caption{Overview of the TED1.8K database}
    \vspace{-1mm}
    \label{tab:database_summary}    
    \renewcommand{\arraystretch}{1.1}
    \begin{tabular}{l  r }
        \hline 
        Total videos & $1864$ \\
        Average duration & $13.7$ min\\   
             \hline \hline
                \multicolumn{2}{c}{YouTube metadata (mean, range)}\\
         \hline
        Views & $247$K, ($40 - 10264$K)\\
        Likes &   $3075$, ($0 - 113$K) \\
        Dislikes & $174$, ($0 - 5750$)\\
        Comments & $462$, ($0 - 26$K)\\
        \hline
    \end{tabular}
        \vspace{-0.05in}
        \label{tab:datasummary}
\end{table}

\section{Proposed Framework}
\label{sec:approach}

\begin{figure*}[tb]
\centering
\includegraphics[width=\textwidth, trim = 0.5cm 0.2cm 0cm 0cm, clip]{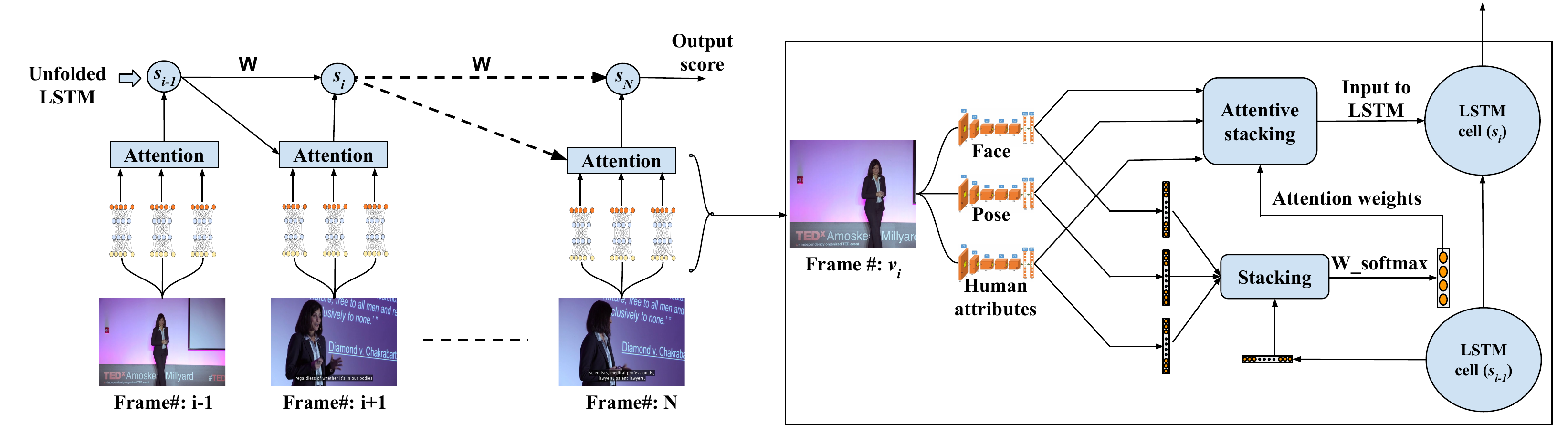}
\caption{Complete architecture of the proposed attention-based LSTM framework.}
\label{fig:horz_arch}
\end{figure*}

In this section, we develop the complete framework for predicting public speaking video popularity from visual cues. Our framework comprises three parts: (i) a collection of independent CNN streams that captures the visual cues, (ii) an attention network that selects the most interesting visual stream at every time step, and (iii) an LSTM network that predicts the popularity scores. Fig.~\ref{fig:horz_arch} shows the complete architecture that we propose in this paper, and below, we describe each part in detail.
\vspace{-0.07in}
\subsection{Encoding visual cues using CNN}
Consider a video $\mathbf{V} = [\mathbf{v}_1, \mathbf{v}_2, \ldots, \mathbf{v}_T]$, where $\mathbf{v}_i$ denotes the $i^{\rm th}$ frame of the video. Each video $\mathbf{V}_j$ is associated with a corresponding video popularity score $y_j\in\R$ computed using the YouTube metadata (see section \ref{sec:database}). In order to visually describe a speaker's presence in a video frame, we propose to extract the following visual cues: physical appearance, gestures and facial expressions. Given a video, we first detect the speaker at every frame by running a state-of-the-art face detector \cite{viola2001rapid}, and a person detector \cite{liu15ssd}. Next, we set up three CNNs which encode the visual appearance and behavior of the speaker in terms of facial and physical attributes, and pose variations.\\

\vspace{-0.08in}
\noindent\textbf{Facial attributes CNN:} The task of this network is to encode facial attributes, such as smile, hairstyle, facial shape, and eye glasses (see Fig.~\ref{fig:teaser}). We expect such
encoding of facial expressions and other attributes to contribute to the popularity of public lecture. This CNN takes the primary face detected in $\mathbf{v}_i$ as an input, and encodes the facial attributes to a descriptor denoted as $\phi(\mathbf{v}_i)$. \\

\vspace{-0.08in}
\noindent\textbf{Pose CNN:} To account for the pose of a speaker, we use another CNN that predicts $17$ landmark points in a human body, e.g.\ knees, elbows (see Fig.~\ref{fig:teaser}). This network encodes the body posture of a speaker in a frame. This feature describes the evolution of the body movements of the speaker, as the
talk goes on. We denote the descriptor as $\theta(\mathbf{v}_i)$ for frame $\mathbf{v}_i$.\\

\vspace{-0.08in}
\noindent\textbf{Physical attributes CNN:} Finally, a third CNN is used to encode the general full body attributes of a speaker, such as, gender, clothing, and age (see Fig.~\ref{fig:teaser}). As noted in psychological studies \cite{riggio1986impression}, the perceived impression of a speaker is likely to be influenced by such physical attributes, and hence we include them as a possible option to attend to by the full network. We denote the full human attribute descriptor as $\psi(\mathbf{v}_i)$ for frame $\mathbf{v}_i$.
\vspace{-0.07in}
\subsection{Multichannel attention network}
\label{subsec:attention}
The next step is to build a multichannel attention network. We propose an attention network that
systematically assigns weights to the three different descriptor channels at every time
step. The computation of weights is based on the content of the current frame, as well as the
information history from all the past frames.   \\

\vspace{-0.08in}
\noindent\textbf{Why attention:} The intuition behind building the attention network is that viewers
do not attend to all the visual cues simultaneously with equal importance. This is even more
relevant in the current set up, where ratings from online viewers' are being considered. It is
possible that due to the variations in camera angle, and editing style, the upper body of a speaker
is not properly visible or detected faces are too small (see Fig.~\ref{fig:attention}). In such
cases, the viewers are likely to rely on the visual cues that are easier to observe. On the other
hand, due to occlusion, low illumination, camera angle, and other factors, the computation of the
frame-level features can also introduce errors. Since our network relies on the history of all
previous frames, an attention-based fusion can help in avoiding propagation of errors.

\indent Given a frame $\mathbf{v}\in \mathbf{V}$, we obtain the three feature channels $\mathbf{f}=\phi(\mathbf{v})$, $\mathbf{p}=\theta(\mathbf{v})$ and 
$\mathbf{c}=\psi(\mathbf{v})$ corresponding to speakers facial attributes, pose, and physical
attributes by doing a forward pass of the respective CNNs. To align the different channels, we pass
them through separate fully connected layers, which maps them to a common output space.
\begin{equation}
\label{eq:alignment}
	\begin{aligned}
	\mathbf{h}_\mathrm{f} & = \tanh(\mathbf{W}_\mathrm{f}\mathbf{f}+ 			\mathbf{b}_\mathrm{f}) \\
	\mathbf{h}_\mathrm{p} & = \tanh(\mathbf{W}_\mathrm{p}\mathbf{p}+ 			\mathbf{b}_\mathrm{p}) \\
	\mathbf{h}_\mathrm{c} & = \tanh(\mathbf{W}_\mathrm{c}\mathbf{c}+ 			\mathbf{b}_\mathrm{c})
	\end{aligned}
\end{equation}
where $\mathbf{W}_\mathrm{f} \in \R^{d_f \times m}, \mathbf{W}_\mathrm{p} \in \R^{d_p\times
m},\mathbf{W}_\mathrm{c} \in \R^{d_c\times m}$ are the projection weights of the three features'
aligned representations, $d_f, d_p, d_c$ are the feature dimensions respectively, $m$ is the
projected feature dimension, and $\mathbf{b}_\mathrm{f}, \mathbf{b}_\mathrm{p}$,
$\mathbf{b}_\mathrm{c}$ are the corresponding biases. The information from all the past frames is
encoded in $\mathbf{h}_\mathrm{s}$ as follows.
\begin{align}
	\centering
	\mathbf{h}_\mathrm{s} &= \tanh(\mathbf{W}_\mathrm{s}\mathbf{s} + \mathbf{b}_\mathrm{s}),
\end{align}
where $\mathbf{s}$ and $\mathbf{W}_\mathrm{s} \in \R^{d_s\times m}$ denote the LSTM states, and the
projection weights for the LSTM states, $d_s$ being the dimension of the LSTM states. The aligned
features, and the history from past frames are then passed through a $2$-layer neural network with a
softmax in the end to generate attention weight distribution over the three channels.
\begin{equation}
\label{eq:attention}
	\begin{aligned}
 \centering
	\mathbf{h}'_\mathrm{j} & = 		\tanh(\mathbf{W}_\mathrm{a}\mathbf{h}_\mathrm{j}+ \mathbf{b}_\mathrm{a}) \,\,\ \forall j \in \{\mathrm{f,p,c,s}\} \\
\mathbf{h}_\mathrm{a} & = [\mathbf{h}'_\mathrm{f}, \mathbf{h}'_\mathrm{p},\mathbf{h}'_\mathrm{c}, \mathbf{h}'_\mathrm{s}]
	\end{aligned}
\end{equation}
\begin{align}
\label{eq:softmax}
\mathbf{a} & = {\rm softmax}(\mathbf{W}_\mathrm{sm}\mathbf{h}_\mathrm{a} + \mathbf{b}_\mathrm{sm})
\end{align}
where $\mathbf{W}_\mathrm{a}\in \R^{m\times n}$ (and $\mathbf{W}_\mathrm{sm} \in \R^{4n\times 3}$ below) are
the weights for the multichannel attention layer. Finally, $\mathbf{a}$ contains the attention
weights corresponding to the facial attributes, pose and physical channels. Once the attention weights are computed, the channel having the maximum attention weight is selected. Let this channel be denoted as $\mathbf{h}^*$, where $\mathbf{h}^*\in \{\mathbf{h}_f, \mathbf{h}_p, \mathbf{h}_c\}$. Ignoring the other channels, only $\mathbf{h}^*$ is input to the LSTM network for regression (described below).

\vspace{-0.07in}
\subsection{LSTM for regression}
As the last part of the proposed architecture, we use a variant of RNN, specifically a
LSTM \cite{hochreiter1997long}, to model the video data as a sequence of frames. LSTMs have been very successful in many sequence modeling tasks with visual
data \cite{wang2016image}. Here we use them to predict video popularity using the multiple visual
cues described above. 

In our architecture, the LSTM cell takes an input $\mathbf{x}_t$ at every time step $t$, and updates
the memory cell $\mathbf{s}_t$ in consultation with its previous state $\mathbf{s}_{t-1}$. At any
time step $t$, $\mathbf{x}_t = \mathbf{h}^*_t$ where $\mathbf{h}^*_t$ is the feature vector with the highest attention at time instant $t$. The LSTM is equipped with several gates which control the update process. A forget gate $\mathsf{g}_t$ decides how much information from the past state $\mathbf{s}_{t-1}$ is carried forward. An input gate $\mathbf{i}_t$ supervises the information from the current input vector $\mathbf{x}_t$. An output gate $\mathbf{o}_t$ puts a check on the information that need to fed to the output as a hidden state. The state update process followed by LSTM is as follows:
\begin{align}
\label{LSTM-eq}
	\begin{aligned}
	\mathbf{g}_t & =  \sigma(\mathbf{W}_g\mathbf{x}_t + \mathbf{W}_g\mathbf{h}_{t-1} + \mathbf{b}_g)\\
    \mathbf{i}_t & = \sigma(\mathbf{W}_i\mathbf{x}_t + \mathbf{W}_i + \mathbf{h}_{t-1} + \mathbf{b}_i)\\
    \mathbf{s}_t & = \mathbf{g}_t\mathbf{s}_{t-1} + \mathbf{i}_t\tanh(\mathbf{W}_s\mathbf{x}_t + \mathbf{W}_s\mathbf{h}_{t-1} + \mathbf{b}_s)\\
    \mathbf{o}_t & = \sigma(\mathbf{W}_o\mathbf{x}_t + \mathbf{W}_o\mathbf{h}_{t-1} + \mathbf{b}_o)\\
	\end{aligned}
\end{align}
\begin{equation}
\centering
\hspace{-11.3em} \mathbf{h}_t = \mathbf{o}_t\tanh(\mathbf{s}_t) 
\end{equation}
where, the $\mathbf{W}$s and $\mathbf{b}$s are the weight matrices and the bias vectors that are learned during the training phase.
\section{Performance Evaluation}
\label{sec:experiments}
In this section, we describe the implementation details, results and comparisons for the proposed multichannel attention-based LSTM network. We perform extensive experiments to evaluate the proposed framework on the TED1.8K corpus (described in Sec.~\ref{sec:database}). We provide insights on how various visual cues contribute to the popularity of a public speaking video. Furthermore, we also analyze the system qualitatively by visualizing how attention is learned in our network.
\vspace{-0.07in}
\subsection{Implementation Details}
The database is randomly divided with ratios $60:20:20$ to create the \train, \val and \test sets. We train the models on the \train set while validating on the \val set, and report the performances on the \test set. We finally report the mean of the performances on multiple runs on different random splits. We plan to make the dataset, the splits and evaluation protocol public, upon publication. \\

\vspace{-0.08in}
\noindent \textbf{Evaluation metric:} We consider two metrics for evaluating the performance of our systems: (i) the Pearson's correlation ($\rho$), and (ii) mean square error (MSE), both computed between the predicted scores and those obtained from YouTube.\\

\vspace{-0.08in}
\noindent \textbf{Facial attributes network:} We use the AlexNet \cite{krizhevsky2012imagenet} CNN for extracting the facial attribute features. The AlexNet is trained on the CelebA database \cite{liu2015faceattributes} to perform facial attribute classification. CelebA is a large facial attributes database consisting a total of $200,000$ images of $10,000$ celebrities, each with $40$ binary attributes. The attributes are diverse, and cover many facial properties, such as `smile', `arched eyebrows', `mouth slightly open', `big lips', `black hair', `oval face', and `mustache'. While training, we initialized the AlexNet with the model parameters pretrained on the ImageNet database \cite{ILSVRC15}. After full training, we obtain a mean class classification accuracy of $79.27\%$ on the CelebA database, which is comparable to the state-of-the-art results reported on the same database \cite{liu2015faceattributes}. Next, we detect a speaker's face in every frame using the Haar cascade face detector \cite{viola2001rapid}. If multiple faces are detected in a frame, we discard the frame due to the ambiguity in determining the speaker's face. We then feedforward the detected (and cropped) face through the trained AlexNet. The $4096$-dimensional output from the last fully connected layer is used as the facial attributes descriptor.\\

\vspace{-0.08in}
\noindent \textbf{Human body detection:} To capture the pose of a speaker in a frame, we employ the
single shot multibox detector (SSD) \cite{liu15ssd}. The SSD is an augmented version of VGGnet \cite{simonyan2014very} trained on the VOC2007 database \cite{pascal-voc-2007} for general object detection, where one of the object categories is `human'. We validated that SSD can recognize the `human' class with an accuracy of $72.5\%$ on VOC2007. Using the SSD detector, we detect the speaker body at every possible frame, and process the cropped part of the frame to further obtain pose and human attributes descriptors.\\

\vspace{-0.08in}
\noindent \textbf{Pose network:} A pretrained VGGnet model \cite{poseactionrcnn} is used to obtain the pose descriptors. This VGGnet is originally trained for keypoint localization and action classification in unconstrained images. The pretrained VGGnet is validated on the VOC2012 \cite{pascal-voc-2012} database that gives an average precision of $70.5\%$ over different action classes. To obtain the pose descriptor, we use the $4096$ dimensional output from the last fully connected layer of this network.\\

\vspace{-0.08in}
\noindent \textbf{Human attributes network:} Similar to the facial attributes features, we use the $4096$ dimensional last fully connected layer
output of the AlexNet to obtain the human attribute features. This AlexNet is trained on a human attributes database \cite{sharma2011learning} with $9344$ human images with $27$ binary attributes. These attributes are related to  physical appearance of a human, such as, `elderly', `wearing t-shirt', `formal men's
suit', `female long skirt', and `standing straight'. The trained model was validated on the human attributes database \cite{sharma2011learning} to yield a mean average precision of $63.7\%$ which is comparable to an earlier report \cite{sharma2011learning}.\\

\vspace{-0.08in}
\noindent \textbf{Alignment and attention:} We downsampled the videos at $5$ frames per second to reduce the amount of data to be processed, primarily due to practical limitations on GPU memory size. We max pool the descriptors (based on ablation experiments detailed below) within a volume of $11$ frames ($\pm 5$ frames around the central frame) and a stride $4$. Thus each volume contains information from $\sim 2$ seconds of video. As described in Section \ref{subsec:attention}, we align the descriptors by projecting them to a common output subspace. We learn $3$ representational layers, one for each descriptor, which projects the respective descriptor to a $1024$ dimensional output space. The aligned descriptors, and the LSTM states from the previous time step are compressed to $128$ dimensions by another fully connected layer, and are stacked together (see Eq.~\eqref{eq:attention}). Finally they are feedforwarded through the last layer, and a \emph{softmax} is applied to obtain attention distribution over different channels (see Eq.~\eqref{eq:softmax}). We use the mean squared error as the loss function, and the RMSProp gradient descent method to learn the parameters of LSTM, and the attention network. Note that the proposed method is trainable end-to-end. However, we have not backpropagated the error into the visual cues networks while training the full system for predicting scores for video popularity.
\begin{table}
\centering
\caption{Performance of SVR in terms of correlation ($\rho$) for different scale and pooling strategies.}
\label{tab:SVR-comparison}
\vspace{-0.1in}
\renewcommand{\arraystretch}{1.1}
	\begin{tabular}{c|c|cccc}
	 \hline
	  	 &  & \multicolumn{4}{|c}{Pooling operation}\\
	 \cline{3-6}
	 & Scale & Mean & Stdev  & Max & Grad\\
	 \hline\hline
	 \multirow{2}{*}{\begin{tabular}[c]{@{}c@{}}Facial attr.\end{tabular}}& single& $0.27$& $0.30$& $0.29$& $0.27$\\
& multi& $0.30$& $0.42$& $\mathbf{0.44}$& $0.39$\\
 \hline
\multirow{2}{*}{\begin{tabular}[c]{@{}c@{}}Pose\end{tabular}}& single& $0.19$& $0.24$& $0.28$& $0.22$\\                                     
& multi& $0.24$& $0.29$& $\mathbf{0.35}$& $0.27$\\ 
 \hline
\multirow{2}{*}{\begin{tabular}[c]{@{}c@{}}Human attr.\end{tabular}} & single& $0.27$& $0.30$& $0.35$& $0.28$\\
& multi& $0.36$& $0.40$& $\mathbf{0.45}$& $0.32$\\ 
 \hline
\end{tabular}
\end{table}
%
\begin{table}
\centering
\caption{Performance of LSTM (obtained on a subset of the original database) in terms of correlation ($\rho$) for frame-level and volume-level features.}
\label{tab:volumetric}
\vspace{-0.1in}
\renewcommand{\arraystretch}{1.1}
\begin{tabular}{c|ccc}
\hline
Features & Frame-level & \begin{tabular}[c]{@{}c@{}}Volume-level \\ multiple pool\end{tabular} & \begin{tabular}[c]{@{}c@{}}Volume-level \\ max pool\end{tabular} \\ \hline \hline
Facial attr. & $0.45$ & $0.51$ & $\mathbf{0.57}$ \\ 
Pose & $0.31$ & $0.39$ & $\mathbf{0.41}$ \\ 
Facial attr + Pose & $0.47$ & $0.51$ & $\mathbf{0.58}$ \\ \hline
\end{tabular}
\end{table}
\vspace{-0.07in}
\subsection{Baselines}
We set up two standard baselines to compare with the proposed approach: (i) support vector regressor (SVR) that uses fixed length vectors as input, and (ii) LSTM (without alignment) that uses time varying vector sequences as inputs. The details of the two baselines are provided below.\\

\vspace{-0.08in}
\noindent \textbf{Support vector regression (SVR):} 
To create video-level feature combining the frame-level features in the video, we use the pooled time series (PoT) representation \cite{ryoo2015pooled}. The PoT scheme considers each dimension of a frame-level feature as a time-series.  Hence, for $d$-dimensional frame-level features, the video consists of $d$ time-series. For each such time series, temporal pooling is performed with a set of temporal filters. We used $4$ pooling operations i.e.\
mean pooling , standard deviation (stdev) pooling, max pooling, and histogram of time series gradient (grad) pooling. Each temporal filter pools over a window with varying size. To achieve this, a temporal pyramid structure \cite{choi2013spatio} is created which incorporate temporal information at $5$ different scales.\\

\vspace{-0.08in}
\noindent\textbf{LSTM regression:}
We set up an LSTM (without alignment) with $50$ hidden nodes for the predicting popularity scores. In our database, the average number of frames per video, after downsampling (@$5$ fps) , is $\sim4300$, which is quite large. Thus we use temporal max pooling within a video volume of $11$ frames with a stride $4$. Thus each volume contains information from $\sim 2$ seconds of video. 
\vspace{-0.08in}
\subsection{Parameter settings}
In this section, we discuss the effects and choice of different parameters used in our framework.\\ 

\vspace{-0.08in}
\noindent \textbf{Pooling operation:}
In order to determine the pooling strategy, we run several prediction experiments using SVR with different pooling operations and temporal window size or scales (see Table \ref{tab:SVR-comparison}). Single scale indicates the window size is equal to the video length, while multi scale creates a temporal pyramid of $5$ levels with varying window size. The results in Table \ref{tab:SVR-comparison} indicate that multiscale max pooling consistently performs better than the other pooling operations. In Table \ref{tab:SVR-comparison}, we have presented results only for the facial attributes, pose, and human attributes channels for brevity. Similar trend was observed for the combinations of channels as well. Following the experimental results, we use \emph{multiscale max pooling} for all our experiments involving both SVR and LSTM.

We also experimented with LSTM-based prediction using features computed (i) at every frame, and (ii) at every volume consisting of $11$ frames. For a given volume, the frame-level features are max pooled to compute the volume-level feature. Table \ref{tab:volumetric} shows the performances of these two types of features, where volume-level features outperform the single frame-based ones.\\

\begin{figure}[t]
\centering
\includegraphics[width=0.90\columnwidth,trim=10 10 0 10, clip]{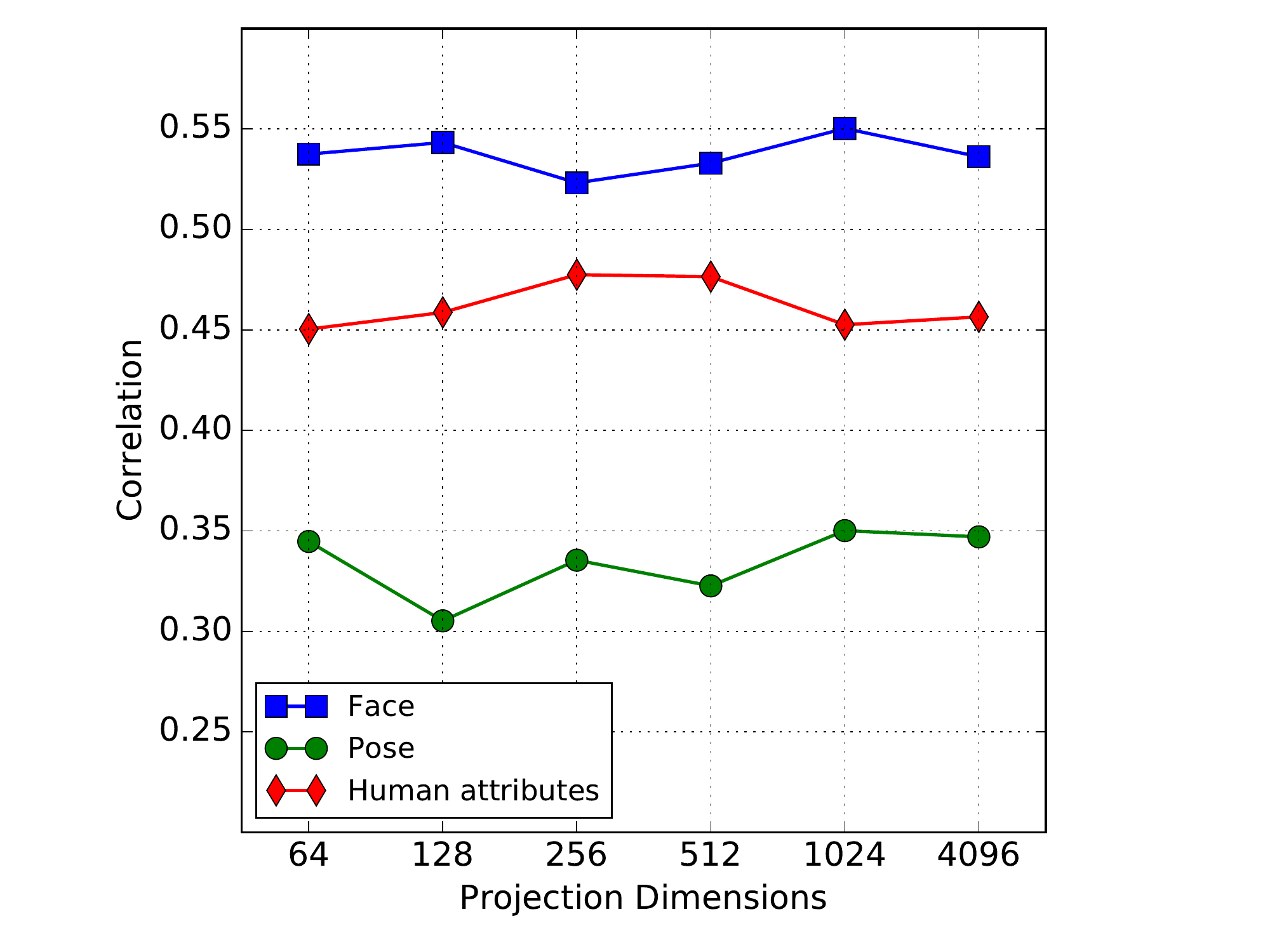}
\caption{Performance variation of individual channel with projection dimension in alignment.}
\label{fig:ablation}
\end{figure}

\vspace{-0.08in}
\noindent \textbf{Projection dimension for alignment:}
Recall that we proposed to align the individual channel to a common output space  (see Section \ref{subsec:attention}). In order to do that initial $4096$-dimensional descriptors are projected onto a smaller dimension. Fig \ref{fig:ablation} shows the performance of each feature channel with different projection dimensions. Fig \ref{fig:ablation} indicates that the $1024$-dimensional representation performs best for all channels except for human attributes. Since our architecture demands the same alignment dimension for all the channels, we chose to use the $1024$.
\vspace{-0.07in}
\subsection{Results}
The performance of the proposed framework for predicting video popularity from visual cues is validated on the TED1.8K database. Our proposed aligned and attention-based LSTM networks are compared against two relevant and challenging baselines, SVR and LSTM. Experiments have been carried out for all possible combinations of the visual channels. Table \ref{tab:corr-main} summarizes the performances of the proposed LSTM networks, and the baselines in terms of two evaluation metrics, correlation $\rho$ and MSE.\\

\vspace{-0.08in}
\noindent
\textbf{SVR vs.\ LSTMs:}
Results in Table \ref{tab:corr-main} show that LSTMs (all variants) perform better than SVR in every case, i.e.\ for each combination of the visual channels. The performance improvement of LSTM is notable when facial attributes and pose channels are used in the network. This results in a significant upgrade in correlation score from $0.45$ (using SVR) to $0.52$ obtained for both the baseline LSTM and the aligned LSTM. The performances follow a similar trend with MSE values as well, where we notice a significant drop in MSE values from $0.87$ to $0.71$ for the case of facial attributes and pose combination. \\

\vspace{-0.08in}
\noindent \textbf{LSTM vs.\ Aligned LSTM:}
The performance of the plain LSTM (baseline), and the proposed aligned LSTM appear to be the same when only correlation scores are considered. However, the MSE values show an improvement for the aligned LSTM, especially when pose channel is included.  This indicates that some improvement in performance could be achieved by using aligned channels, as compared to using original features in the LSTM. Both variants perform the best for the combination involving facial attributes and pose variations yielding a correlation score of $0.52$ in both the cases.\\

\vspace{-0.08in}
\noindent \textbf{Multichannel attention LSTM:}
The proposed temporal attention-based LSTM network that selects only one (the most relevant) visual channel at every time step
outperforms SVR in all cases, and other LSTMs in most cases in terms of both $\rho$ and MSE. We also experimented with using weighted versions of all three channels, instead of choosing the one with highest attention weight. This result was not superior to our current scheme of selecting only the best channel at a time step. From Table \ref{tab:corr-main} it is clear that unlike LSTM or aligned LSTM, the performance of the attention network stands out when all three channels are added, for it achieves a correlation score of $0.57$ - the best observed value in all possible scenarios. There is also a significant drop in corresponding MSE values of attention-based LSTM, when using all three channels. This supports our hypothesis  that attention is an important contributor to how the visual attributes of a public speaker is perceived.\\
\begin{table}
\centering
\caption{Performance of the proposed framework on the TED1.8K database \scriptsize{(Face: facial attributes, HAT: human attributes).}}
\label{tab:corr-main}
\begin{tabular}{l| c c c c}
\hline
Method & \multicolumn{1}{c}{\begin{tabular}[c]{@{}c@{}}Pose +\\ HAT \end{tabular}} & \multicolumn{1}{c}{\begin{tabular}[c]{@{}c@{}}Face + \\ HAT \end{tabular}} & \multicolumn{1}{c}{\begin{tabular}[c]{@{}c@{}}Face +\\  Pose\end{tabular}} & \multicolumn{1}{c}{\begin{tabular}[c]{@{}c@{}}Face + Pose\\ + HAT\end{tabular}} \\ \hline \hline
& \multicolumn{4}{c}{Correlation ($\rho$)}\\
\cline{1-5}
SVR (baseline) & $0.44$ & $0.47$ & $0.46$ & 0.48 \\ 
LSTM (baseline) & $0.45$ & $0.51$ & $0.52$ & $0.51$\\ 
Aligned LSTM & $\mathbf{0.45}$ & $0.51$& $0.52$& $0.51$\\ 
\bf Attention LSTM & $0.41$ & $\mathbf{0.51}$& $\mathbf{0.55}$ & $\mathbf{0.57}$ \\ \hline
& \multicolumn{4}{c}{MSE} \\ \hline
SVR (baseline) &  $0.87$ & $0.84$ & $0.87$ & $0.76$ \\ 
LSTM (baseline) & $0.81$ & $0.72$ & $0.71$ & $0.73$ \\ 
Aligned LSTM & $\mathbf{0.74}$ & $\mathbf{0.70}$ & $0.69$ & $0.76$ \\ 
\bf Attention LSTM & $0.80$ & $0.74$ & $\bf{0.68}$ & $\bf{0.68}$ \\ \hline
\end{tabular}
\end{table}
\begin{figure*}[tb]
\centering
\begin{minipage}{\linewidth}
	\setlength{\fboxsep}{0 pt}%
	\setlength{\fboxrule}{2pt}%
	\fbox{\includegraphics[width=\textwidth, trim = 0cm 0cm 0cm 0cm]{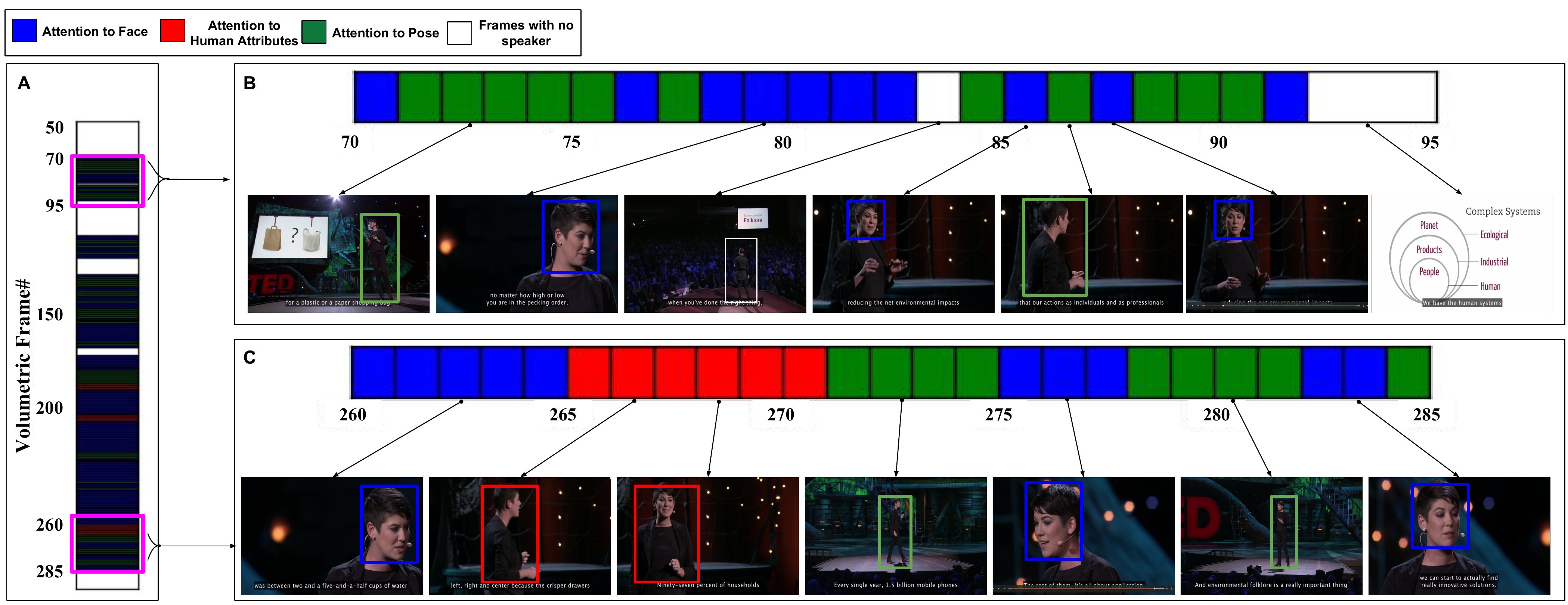}}\\
	\vspace{0.1in}
\end{minipage}
\begin{minipage}{\linewidth}
	\setlength{\fboxsep}{0pt}%
	\setlength{\fboxrule}{2pt}%
	\fbox{\includegraphics[width=\textwidth, trim = 0cm 0cm 0cm 0cm]{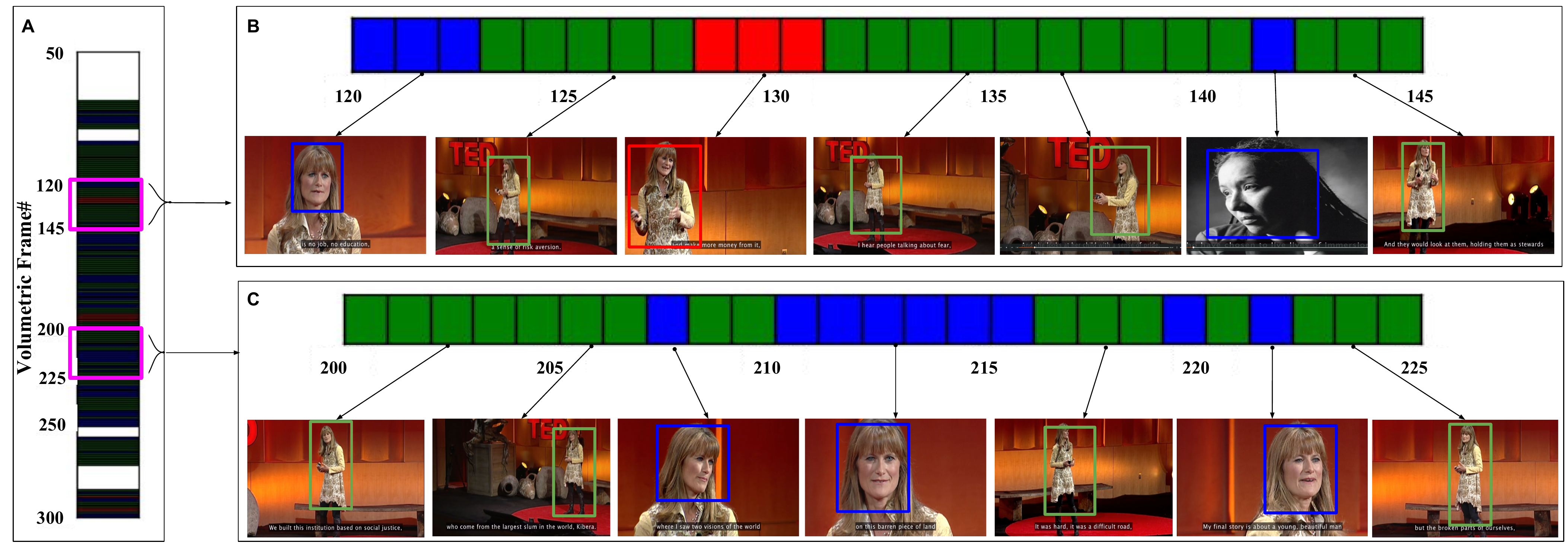}}
\end{minipage}
\vspace{-1em}
\caption{Visualization of attention across the visual channels in two videos, as learned by our attention-based LSTM network.}
\label{fig:attention}
\end{figure*}

\vspace{-0.08in}
\noindent\textbf{Visualizing attention:}
In addition to the quantitative results, we are also interested in gaining further insights into the operation of the attention-based network. We thus look into the attention weights learned by the network at every time step to see if it corresponds with our intuitive understanding of attention to visual cues. Fig.~\ref{fig:attention} presents visualization results of the attention network for two sample videos from TED1.8K database. The figure shows the sequence of the chosen visual channel (corresponding to the maximum attention weight) for each time step as generated by our network. This is computed at volume-level 
(comprises $11$ frames) of the video as described in earlier sections. In Fig.~\ref{fig:attention}, block \texttt{A} shows which channel has the highest attention at every time step (volume) of the video, while block \texttt{B} and \texttt{C} expand specific parts of the video to demonstrate the details of how attention works in our system. A representative frame from the corresponding part of the video is shown for easy validation. It is clear that when the speaker's face is shown on screen up-close, our attention network correctly selects the facial attributes channel as the most important cue. Likewise, when the full upper body is visible, attention switches to the human attributes; and when the speaker is at a
distance, the pose channel is selected by the attention network. Also note that the frames with no
speaker present in them are also correctly rejected by our system (white blocks). This visualization of the learned
attention weights increases interpretability, and also aligns with our intuitive human understanding
and perception.
\vspace{-0.07in}
\subsection{Discussion}
From the various experimental results, we observe that human attributes and pose together form the weakest combination of all visual cues. However, when the pose or human attributes channel is combined with the facial attributes channel prediction accuracy improves. This could be intuitively explained by the fact that during TED talks, closeup of the speakers' faces appear on-screen more frequently, and perhaps influence the viewers' perception more than other cues. Hence combining facial attributes with any other visual cues provide better results. In other words, facial attributes are the most important among the visual cues we considered.

We also notice that the our attention-based LSTM performs the best when all three channels are included. This suggests that attention-based networks are particularly useful for dealing with larger number of channels, where it is natural to switch attention from one cue to another. We expect that our attention network will be further useful when the attention is sought among multiple modalities, say, visual, speech and lexical channels.

The visualization in Fig.~\ref{fig:attention} shows that the duration for which our system selects the human
attributes channel (indicated by the red blocks) is shorter compared to the other two channels. This also aligns with our previous observation that the human attributes channel may be weaker than the others as these attributes do not change much over the duration of the video. This can also be seen from Table \ref{tab:corr-main}, where improvement in $\rho$ between the face-pose and face-pose-human attributes is not very large, while MSE values remain the same.

A limitation of our framework lies in detecting the presence of a speaker in a video. As seen in Fig.~\ref{fig:attention}, our system can correctly ignore the frames where there is no speaker present. However, it also rejects the frames where more than one person are detected, e.g.\ when faces are detected in the audience along with the speaker's face. Our current system also can not distinguish between a speaker and any other human face shown in slides or in audience. This can be observed in the second example in Fig.~\ref{fig:attention}-block \texttt{B}.
\section{Conclusion}
\label{sec:conclusion}
In this work, we have analyzed the visual behavior of speakers in the context of public speaking. To this end, we proposed a computational framework for predicting popularity of public speaking videos from visual cues related to physical appearance, facial expressions, and pose variations (extracted using pretrained CNN models). Our framework selects the visual channel with maximum attention (learned as a latent variable) at every time step, and uses the sequence of the selected channels to predict video popularity. Note that the full architecture is trainable end-to-end; however, since the CNN streams require large amount of data to train, they are initialized with networks pretrained on appropriate data with the respective sources being different. 

The proposed multichannel attention network can learn a human-like selective attention mechanism depending on the visual cues present on screen. A large database comprising 1864 TED talk videos and the corresponding YouTube metadata was constructed to facilitate the current study, and the study of public speaking in general. Extensive experiments on this database showed that our framework can predict video popularity from visual cues alone with significant accuracy outperforming challenging baselines. We observe that facial attributes contribute towards video popularity more than the other visual cues, while the human attributes (HAT) channel gets attended fewer times in the network.

Our proposed framework can be easily extended to include acoustic and lexical channels if prediction of video popularity is the primary objective. The prediction accuracy is also expected to improve significantly if verbal cues (e.g.~speech prosody, intonation, fluency) are added to our framework. 
\section*{Acknowledgement}
\label{sec:conclusion}
The authors would like to thank NVIDIA Corp.~for donating a TitanX GPU for supporting this research.

\balance

\ifCLASSOPTIONcaptionsoff
  \newpage
\fi

\bibliographystyle{IEEEtran}
\bibliography{IEEEabrv,sigproc,egbib} 

\begin{thebibliography}{10}
\providecommand{\url}[1]{#1}
\csname url@samestyle\endcsname
\providecommand{\newblock}{\relax}
\providecommand{\bibinfo}[2]{#2}
\providecommand{\BIBentrySTDinterwordspacing}{\spaceskip=0pt\relax}
\providecommand{\BIBentryALTinterwordstretchfactor}{4}
\providecommand{\BIBentryALTinterwordspacing}{\spaceskip=\fontdimen2\font plus
\BIBentryALTinterwordstretchfactor\fontdimen3\font minus
  \fontdimen4\font\relax}
\providecommand{\BIBforeignlanguage}[2]{{%
\expandafter\ifx\csname l@#1\endcsname\relax
\typeout{** WARNING: IEEEtran.bst: No hyphenation pattern has been}%
\typeout{** loaded for the language `#1'. Using the pattern for}%
\typeout{** the default language instead.}%
\else
\language=\csname l@#1\endcsname
\fi
#2}}
\providecommand{\BIBdecl}{\relax}
\BIBdecl

\bibitem{vinciarelli2009social}
A.~Vinciarelli, M.~Pantic, and H.~Bourlard, ``Social signal processing: Survey
  of an emerging domain,'' \emph{Image and Vision Computing}, vol.~27, no.~12,
  pp. 1743--1759, 2009.

\bibitem{xiao2015head}
B.~Xiao, P.~Georgiou, B.~Baucom, and S.~S. Narayanan, ``Head motion modeling
  for human behavior analysis in dyadic interaction,'' \emph{IEEE transactions
  on multimedia}, vol.~17, no.~7, pp. 1107--1119, 2015.

\bibitem{batrinca2016multimodal}
L.~Batrinca, N.~Mana, B.~Lepri, N.~Sebe, and F.~Pianesi, ``Multimodal
  personality recognition in collaborative goal-oriented tasks,'' \emph{IEEE
  Transactions on Multimedia}, vol.~18, no.~4, pp. 659--673, 2016.

\bibitem{aran2014broadcasting}
O.~Aran, J.-I. Biel, and D.~Gatica-Perez, ``Broadcasting oneself: Visual
  discovery of vlogging styles,'' \emph{IEEE Transactions on multimedia},
  vol.~16, no.~1, pp. 201--215, 2014.

\bibitem{biel2013youtube}
J.-I. Biel and D.~Gatica-Perez, ``The youtube lens: Crowdsourced personality
  impressions and audiovisual analysis of vlogs,'' \emph{IEEE Transactions on
  Multimedia}, vol.~15, no.~1, pp. 41--55, 2013.

\bibitem{wesson2005communication}
C.~J. Wesson, ``The communication and influence of confidence and
  uncertainty,'' Ph.D. dissertation, University of Wolverhampton, UK, 2005.

\bibitem{riggio1986impression}
R.~E. Riggio and H.~S. Friedman, ``Impression formation: The role of expressive
  behavior.'' \emph{Journal of Personality and Social Psychology}, vol.~50,
  no.~2, p. 421, 1986.

\bibitem{cho2015describing}
K.~Cho, A.~Courville, and Y.~Bengio, ``Describing multimedia content using
  attention-based encoder-decoder networks,'' \emph{IEEE Transactions on
  Multimedia}, vol.~17, no.~11, pp. 1875--1886, 2015.

\bibitem{yang2016stacked}
Z.~Yang, X.~He, J.~Gao, L.~Deng, and A.~Smola, ``Stacked attention networks for
  image question answering,'' in \emph{Proceedings of the IEEE Conference on
  Computer Vision and Pattern Recognition}, 2016.

\bibitem{albert1970physical}
S.~Albert and J.~M. Dabbs~Jr, ``Physical distance and persuasion.''
  \emph{Journal of personality and social psychology}, vol.~15, no.~3, p. 265,
  1970.

\bibitem{pertaub2002experiment}
D.~P. Pertaub, M.~Slater, and C.~Barker, ``An experiment on public speaking
  anxiety in response to three different types of virtual audience,''
  \emph{Presence: Teleoperators and virtual environments}, vol.~11, no.~1, pp.
  68--78, 2002.

\bibitem{daly1989self}
J.~A. Daly, A.~L. Vangelisti, and S.~G. Lawrence, ``Self-focused attention and
  public speaking anxiety,'' \emph{Personality and Individual Differences},
  vol.~10, no.~8, pp. 903--913, 1989.

\bibitem{rosenberg2005acoustic}
A.~Rosenberg and J.~Hirschberg, ``Acoustic/prosodic and lexical correlates of
  charismatic speech.'' in \emph{INTERSPEECH}, 2005, pp. 513--516.

\bibitem{strangert2008makes}
E.~Strangert and J.~Gustafson, ``What makes a good speaker? subject ratings,
  acoustic measurements and perceptual evaluations.'' in \emph{INTERSPEECH},
  vol.~8, 2008, pp. 1688--1691.

\bibitem{gonzalez2013perceptually}
S.~Gonzalez and X.~Anguera, ``Perceptually inspired features for speaker
  likability classification.'' in \emph{ICASSP}, 2013, pp. 8490--8494.

\bibitem{tanveer2015unsupervised}
M.~I. Tanveer, J.~Liu, and M.~E. Hoque, ``Unsupervised extraction of
  human-interpretable nonverbal behavioral cues in a public speaking
  scenario,'' in \emph{Proceedings of the 23rd ACM international conference on
  Multimedia}.\hskip 1em plus 0.5em minus 0.4em\relax ACM, 2015, pp. 863--866.

\bibitem{wortwein2015multimodal}
T.~W{\"o}rtwein, M.~Chollet, B.~Schauerte, L.~P. Morency, R.~Stiefelhagen, and
  S.~Scherer, ``Multimodal public speaking performance assessment,'' in
  \emph{Proceedings of the 2015 ACM on International Conference on Multimodal
  Interaction}, 2015, pp. 43--50.

\bibitem{wortwein2015automatic}
T.~W{\"o}rtwein, L.~P. Morency, and S.~Scherer, ``Automatic assessment and
  analysis of public speaking anxiety: A virtual audience case study,'' in
  \emph{Affective Computing and Intelligent Interaction (ACII), 2015
  International Conference on}, 2015, pp. 187--193.

\bibitem{nguyen2013multimodal}
L.~S. Nguyen, A.~Marcos-Ramiro, M.~Marr{\'o}n~Romera, and D.~Gatica-Perez,
  ``Multimodal analysis of body communication cues in employment interviews,''
  in \emph{Proceedings of the 15th ACM on International conference on
  multimodal interaction}.\hskip 1em plus 0.5em minus 0.4em\relax ACM, 2013,
  pp. 437--444.

\bibitem{nojavanasghari2016deep}
B.~Nojavanasghari, D.~Gopinath, J.~Koushik, T.~Baltru{\v{s}}aitis, and L.~P.
  Morency, ``Deep multimodal fusion for persuasiveness prediction,'' in
  \emph{Proceedings of the 18th ACM International Conference on Multimodal
  Interaction}.\hskip 1em plus 0.5em minus 0.4em\relax ACM, 2016, pp. 284--288.

\bibitem{krizhevsky2012imagenet}
A.~Krizhevsky, I.~Sutskever, and G.~E. Hinton, ``Imagenet classification with
  deep convolutional neural networks,'' in \emph{Advances in neural information
  processing systems}, 2012, pp. 1097--1105.

\bibitem{simonyan2014very}
K.~Simonyan and A.~Zisserman, ``Very deep convolutional networks for
  large-scale image recognition,'' \emph{arXiv preprint arXiv:1409.1556}, 2014.

\bibitem{hochreiter1997long}
S.~Hochreiter and J.~Schmidhuber, ``Long short-term memory,'' \emph{Neural
  computation}, vol.~9, no.~8, pp. 1735--1780, 1997.

\bibitem{adascan2017}
A.~Kar, N.~Rai, K.~Sikka, and G.~Sharma, ``Adascan: Adaptive scan pooling in
  deep convolutional neural networks for human action recognition in videos.''

\bibitem{viola2001rapid}
P.~Viola and M.~Jones, ``Rapid object detection using a boosted cascade of
  simple features,'' in \emph{Computer Vision and Pattern Recognition, 2001.
  CVPR 2001. Proceedings of the 2001 IEEE Computer Society Conference on},
  vol.~1.\hskip 1em plus 0.5em minus 0.4em\relax IEEE, 2001, pp. I--I.

\bibitem{liu15ssd}
W.~Liu, D.~Anguelov, D.~Erhan, C.~Szegedy, S.~Reed, C.-Y. Fu, and A.~C. Berg,
  ``{SSD}: Single shot multibox detector,'' \emph{arXiv preprint
  arXiv:1512.02325}, 2015.

\bibitem{wang2016image}
C.~Wang, H.~Yang, C.~Bartz, and C.~Meinel, ``Image captioning with deep
  bidirectional lstms,'' in \emph{ACM Multimedia}.\hskip 1em plus 0.5em minus
  0.4em\relax ACM, 2016, pp. 988--997.

\bibitem{liu2015faceattributes}
Z.~Liu, P.~Luo, X.~Wang, and X.~Tang, ``Deep learning face attributes in the
  wild,'' in \emph{Proceedings of International Conference on Computer Vision
  (ICCV)}, 2015.

\bibitem{ILSVRC15}
O.~Russakovsky, J.~Deng, H.~Su, J.~Krause, S.~Satheesh, S.~Ma, Z.~Huang,
  A.~Karpathy, A.~Khosla, M.~Bernstein, A.~C. Berg, and L.~Fei-Fei, ``{ImageNet
  Large Scale Visual Recognition Challenge},'' \emph{International Journal of
  Computer Vision (IJCV)}, vol. 115, no.~3, pp. 211--252, 2015.

\bibitem{pascal-voc-2007}
\BIBentryALTinterwordspacing
M.~Everingham, L.~Van~Gool, C.~K.~I. Williams, J.~Winn, and A.~Zisserman, ``The
  {PASCAL} {V}isual {O}bject {C}lasses {C}hallenge 2007 {(VOC2007)}
  {R}esults,'' 2007. [Online]. Available:
  \url{http://www.pascal-network.org/challenges/VOC/voc2007/workshop/index.html}
\BIBentrySTDinterwordspacing

\bibitem{poseactionrcnn}
G.~Gkioxari, B.~Hariharan, R.~Girshick, and J.~Malik, ``R-cnns for pose
  estimation and action detection,'' 2014.

\bibitem{pascal-voc-2012}
M.~Everingham, L.~Van~Gool, C.~K.~I. Williams, J.~Winn, and A.~Zisserman, ``The
  {PASCAL} {V}isual {O}bject {C}lasses {C}hallenge 2012 {(VOC2012)}
  {R}esults,''
  http://www.pascal-network.org/challenges/VOC/voc2012/workshop/index.html.

\bibitem{sharma2011learning}
G.~Sharma and F.~Jurie, ``Learning discriminative spatial representation for
  image classification,'' in \emph{BMVC 2011-British Machine Vision
  Conference}.\hskip 1em plus 0.5em minus 0.4em\relax BMVA Press, 2011, pp.
  1--11.

\bibitem{ryoo2015pooled}
M.~S. Ryoo, B.~Rothrock, and L.~Matthies, ``Pooled motion features for
  first-person videos,'' in \emph{Proceedings of the IEEE Conference on
  Computer Vision and Pattern Recognition}, 2015, pp. 896--904.

\bibitem{choi2013spatio}
J.~Choi, Z.~Wang, S.~C. Lee, and W.~J. Jeon, ``A spatio-temporal pyramid
  matching for video retrieval,'' \emph{Computer Vision and Image
  Understanding}, vol. 117, no.~6, pp. 660--669, 2013.

\end{thebibliography}
\end{document}